\documentclass[showkeys,showpacs,superscriptaddress,preprintnumbers,byrevtex,nofootinbib]{revtex4}
\usepackage{amsmath,amsfonts,amssymb,amscd,amsxtra,amsthm}
\usepackage{graphicx}
\usepackage{bm}
\usepackage{epstopdf}
\begin{document}
\preprint{KIAS-P11055}
\title{Photoproduction of $K^*$ meson off the proton target with the
Regge contributions}
%-------------------------------------------------
\author{Seung-il Nam}
\email[E-mail: ]{sinam@kias.re.kr}
\affiliation{School of Physics, Korea Institute for Advanced Study (KIAS), Seoul 130-722, Republic
of Korea}
%-------------------------------------------------
\author{Byung-Geel Yu}
\email[E-mail: ]{bgyu@kau.ac.kr}
\affiliation{Research Institute for Basic Sciences, Korea Aerospace
University, Goyang 412-791, Republic of Korea}
%--------------------------------------------------
\date{\today}
\begin{abstract}
We investigate the $K^*\Lambda(1116)$  photoproduction off the proton target, employing the tree-level Born approximation within the effective Lagrangian approach. We take into account the $\kappa$, $K$, and $K^*$ exchanges in the $t$ channel in addition to the ground-state nucleon- and hyperon-pole contributions in the $s$ and $u$ channels with the gauge-invariant form-factor scheme. The nucleon and hyperon resonances, $D_{13}(2080)$ and $\Sigma^*(1385)$ are also included. We take into account the Regge trajectories of the strange-mesons for the $t$-channel contributions in such a way that the Feynman and Regge propagators are interpolated smoothly for the relevant photon energy region. We provide numerical results for the differential cross sections, which result in qualitatively good agreement with the presently available experimental data from CLAS collaboration at Jefferson laboratory. It turns out that the Feynman-Regge interpolation for the meson exchanges play a critical role in reproducing the data simultaneously for the low and relatively high energy regions. We also compute the total cross sections, momentum-transfer dependence of the differential cross sections in the $t$-channel, and the photon-beam asymmetry as useful guides for future experiments.
\end{abstract}
\pacs{12.38.Lg, 14.40.Aq} \keywords{$K^*\Lambda$ photoproduction,
Regge trajectory, strangeness production, magnetic-dipole moments}
\maketitle
%--------------------------------------------------
\section{Introduction}
%--------------------------------------------------
Strange-meson  photoproduction is one of the important subjects in hadron physics for decades. Employing the structureless-clean probe in the relatively-low energy region, i.e. the photon beam with the nucleon target, one can understand the production mechanisms for the strange hadrons and obtain various information on their internal structures, beyond the SU(2) isospin symmetry, although the photon behaves as a structural particle at very high-energy regions, which are far beyond our scope. Among various strangeness production processes, the kaon photoproduction has been investigated intensively in both sides of experiment~\cite{Bradford:2005pt,Achenbach:2011rf,Hicks:2010pg,Tsukada:2007jy,Watanabe:2006bs} and theory~\cite{Janssen:2001wk,Mart:2011ez,Yu:2011fv,DeCruz:2010ry}, which indicate a considerable amount of resonant contributions in addition to the non-resonant background ones. In contrast to that, the strange vector meson, $K^*(892,1^-)$ had not been well explored to date experimentally~\cite{Hleiqawi:2005sz,Hicks:2008yn,Hleiqawi:2007ad}, whereas there have been several theoretical works~\cite{Zhao:2001jw,Nam:2005jb,Oh:2006hm,Ozaki:2009wp,Oh:2006in,Thornber:1968mw}. However,  it becomes possible to measure the observables in
photoproduction of the heavier meson with much more statistics, as the photon-beam experimental facilities are recently planning to upgrade their photon-beam energy, such as LEPS2 at SPring-8 and CLAS12 at Jefferson laboratory~\cite{Yosoi:2011zz,Salgado:2011zz,Livingston:2011zz}. Therefore, it is an urgent and important task to provide theoretical predictions as a guide for these new experiments. We note that there are several experimental data and ongoing analyses for the $K^*$ photoproduction by LEPS~\cite{Hwang2012} and CLAS~\cite{Hicks2012} collaborations.

In the present work,  we would like to investigate the $K^*$ photoproduction from the proton target, $\gamma N\to K^*\Lambda(1116)$. As a theoretical framework, we employ the effective Lagrangian approach at the tree-level Born approximation. We also make use of the gauge-invariant prescription for the form factors which conserves the photoproduction current in the presence of them, satisfying the Ward-Takahashi identity for the scattering amplitude~\cite{Haberzettl:1998eq,Davidson:2001rk,Haberzettl:2006bn}. This phenomenological prescription for form factors also fulfills the crossing symmetry and on-shell condition, simultaneously. We already have verified that this prescription reproduces the experimental data qualitatively well for various hyperon photoproductions~\cite{Nam:2005jb,Nam:2005uq,Nam:2009cv,Nam:2010au,Nam:2011np}. At the tree level, we will take into account the baryon-pole contributions from the ground-state proton and hyperons ($\Lambda$, $\Sigma$, $\Sigma^*$) in the $s$ and $u$ channels, respectively.  In addition, we consider the hyperon and nucleon resonances, $\Sigma^*(1385)$ and $D_{13}(2080)$. We notice that $D_{13}$ was turned out to be the most crucial contribution in the present reaction process~\cite{Kim:2011rm}. The strange meson-pole contributions are also considered, using the scalar, pseudoscalar, and vector kaon exchanges, i.e. $\kappa$, $K$, and $K^*$, respectively. As in Refs.~\cite{Regge:1959mz,Vanderhaeghen:1997ts,Corthals:2005ce,Corthals:2006nz,Ozaki:2009wp}, the Regge trajectories in the $t$ channel play an important role in reproducing experimental data especially in the high-energy region. However, as discussed in Refs.~\cite{Nam:2010au,Nam:2011np}, even in the relatively low-energy region, the Regge trajectories contribute to a certain extend, resulting in the possible smooth interpolation between the Feynman and Regge propagators in the intermediate-energy region. Hence, following the idea of Refs.~\cite{Nam:2010au,Nam:2011np}, we make use of this interpolation prescription in a simplified form. We also perform the estimations for the theoretical uncertainties in the numerical results, according to the various parameters in the present work. 

We provide  numerical results for the angular and energy dependences of the cross sections for the present reaction processes, i.e. $d\sigma/d\cos\theta$ and $\sigma$. The numerical results show qualitatively good agreement with the presently available but preliminary experimental data for the angular dependences from CLAS collaboration at Jefferson laboratory~\footnote{We note the experimental data were approved to be used in this work by the author of Ref.~\cite{Hicks:2010pg} as indicated in the acknowledgment.}~\cite{Hicks:2010pg}. We observe that the contribution of the magnetic-dipole moment of the $K^*$ from the $\gamma K^* K^*$ vertex is crucial to reproduce the data especially in the forward-scattering region. It turns out that the interpolation of the Feynman-Regge prescription plays a crucial role in reproducing the data qualitatively well for the low and relatively high energy regions simultaneously. The $D_{13}$ contribution performs also a significant role in producing the total cross section. Moreover, we also present the numerical results for the $t$-channel momentum-transfer dependence, $d\sigma/dt$ and the photon-beam asymmetry $\Sigma$ for future experiments.

The present  work is structured as follows. In Section II, we briefly introduce the theoretical framework for the computation of the $K^*\Lambda$ photoproduction off the proton target. The numerical results with related discussions are given in Section III. We close the present work with summary and conclusion in Section IV.

%--------------------------------------------------
\section{Theoretical Formalisms}
%--------------------------------------------------
%FIGURE>>>
\begin{figure}[t]
\includegraphics[width=12cm]{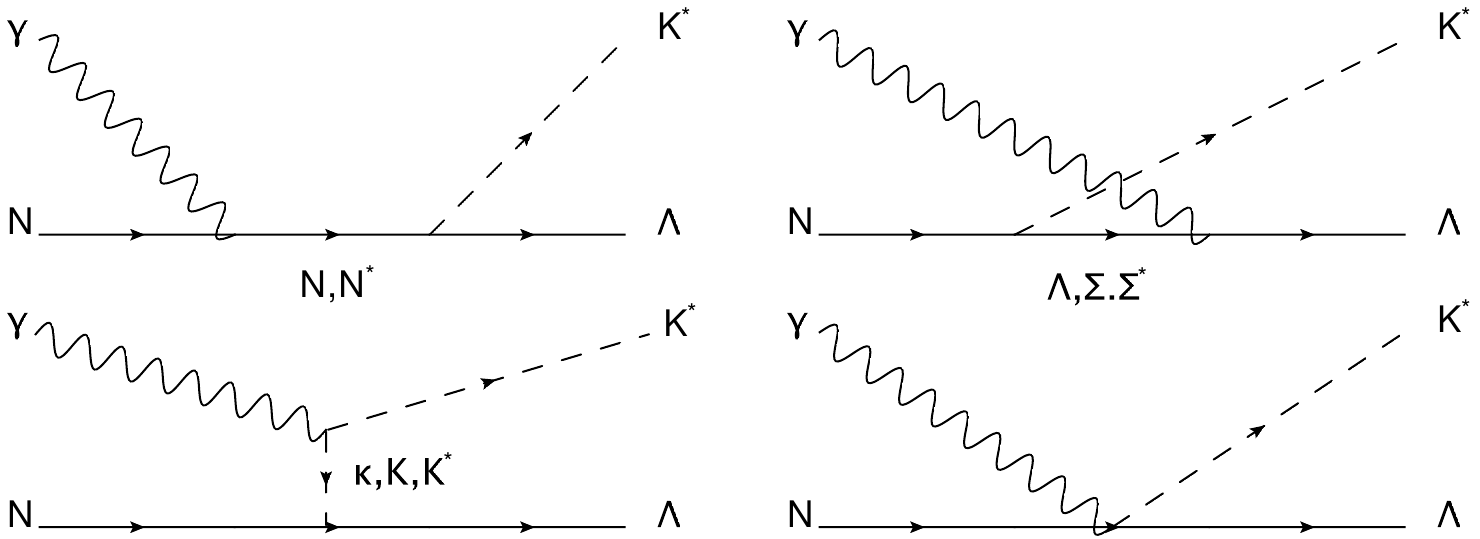}
\caption{Relevant Feynman diagrams for $K^{*}$ photoproduction off the nucleon target at tree level. Four momenta of the incident photon, target nucleon, outgoing $K^{*}$, and recoil hyperon are denoted by $k_{1}$, $p_{1}$, $k_{2}$, and $p_{2}$, respectively. Ground-state and resonant nucleon- and hyperon-pole contributions are depicted in the $s$- and $u$-channel, respectively, whereas those of strangeness meson-poles are in the $t$ channel. The contact term is necessary for gauge invariance of the photoproduction amplitude.}
\label{FIG0}
\end{figure}
%FIGURE<<<

In  this Section, we explain briefly the theoretical framework to compute the $K^*\Lambda$ photoproduction off the nucleon target at tree level in general. The relevant Feynman diagrams for the present production process are depicted in Figure~\ref{FIG0}, where the four momenta of the particles are defined. We take into account $(N,N^*)$-nucleon and $(\Lambda,\Sigma,\Sigma^*)
$-hyperon poles in the $s$ and $u$ channels, respectively. In the $t$ channel, there are three different strange-meson exchanges, i.e. those from the scalar $\kappa(800,0^+)$, pseudoscalar $K(495,0^-)$, and vector $K^*(892,1^-)$ mesons. In addition to these conventional Born
contributions, one needs the contact-interaction term for gauge invariance of the scattering amplitude in terms of the Ward-Takahashi (WT) identity. Now, we first define the Lagrangians for all the electromagnetic (EM) interaction vertices as follows:
%EQUATION>>>
\begin{eqnarray}
\label{eq:LAGEM}
\mathcal{L}_{\gamma BB}&=&-\bar{B}\left[e_B\rlap{\,/}{A}
-\frac{e \kappa_B}{4M_B}(\sigma\cdot F)\right]B,
%-----------------------
\cr
\mathcal{L}_{\gamma PP}&=&e_P\left[P(\partial_\mu P^{\dagger})
-P^{\dagger}(\partial_\mu P) \right]A^\mu,
%-----------------------
\cr
\mathcal{L}_{\gamma VV}&=&ie_{V}\left[V^{\nu}G^{\dagger}_{\mu\nu}
-V^{\dagger\nu}G_{\mu\nu}\right]A^\mu
+ie \mu_VV^{\dagger\mu}V^\nu
F_{\mu\nu}
+\frac{ie \mathcal{Q}_V}{M^2_V}
G^{\dagger}_{\mu\nu}G_{\rho}^{\nu}F^{\rho\mu},
\cr
%-----------------------
\mathcal{L}_{\gamma PV}&=&g_{\gamma PV}\epsilon_{\mu\nu\sigma\rho}
(\partial_\mu A_\nu)(\partial_\sigma V_\rho)P+\mathrm{h.c.},
\cr
%-----------------------
\mathcal{L}_{\gamma SV}&=&g_{\gamma SV}SG_{\mu\nu}F^{\mu\nu}
+\mathrm{h.c.},
%-----------------------
\end{eqnarray}
%EQUAITON<<<
where  $B$, $A_\mu$, $P$, $V$, and $S$ stand for the spin-$1/2$
baryon, photon, pseudoscalar $K(495,0^-)$, vector $K^*(892,1^-)$,
and scalar $\kappa(800,0^+)$ strange-meson fields, respectively. The $e_h$ and $e$ for the electric charge
for a hadron $h$ and unit electric charge. Similarly, we set $M_h$
to be the mass of the hadron $h$. The field strength tensors for
the photon and $K^*$ are assigned as $F_{\mu\nu}$ and
$G_{\mu\nu}$, and the $\sigma\cdot F$ stands for
$\sigma_{\mu\nu}F^{\mu\nu}$, where the antisymmetric tensor is
given by $\sigma_{\mu\nu}\equiv
i(\gamma_\mu\gamma_\nu-\gamma_\nu\gamma_\mu)/2$. The nucleon
anomalous magnetic moment is denoted by $\kappa_N$, employing
$\kappa_p=1.79$ for the proton for
instance~\cite{Nakamura:2010zzi}. We also use the following values
for the hyperon electromagnetic (EM) vertices:
$\kappa_\Lambda=-0.61$ for $\gamma\Lambda\Lambda$ and
$\kappa_{\Lambda\Sigma^*}=1.61$ for $\gamma\Lambda\Sigma^{*}$,
respectively~\cite{Nakamura:2010zzi}. The mesonic
magnetic-transition couplings can be computed employing the
experimental information~\cite{Nakamura:2010zzi} and
estimated as:
%EQUATION>>>
\begin{equation}
\label{eq:GGGG} g^{(\mathrm{charged},\mathrm{neutral})}_{\gamma
KK^*}= (-0.254/\mathrm{GeV},-0.388/\mathrm{GeV}),\,\,\,\,\,
g^{(\mathrm{charged},\mathrm{neutral})}_{\gamma \kappa K^*}=
(0.120/\mathrm{GeV},0.240/\mathrm{GeV}).
\end{equation}
%EQUAITON<<<
We note that, in consideration of the CP-conserving $\gamma VV$ vertex, it contains three different EM couplings as the electric-monopole ($e_V$), magnetic-dipole ($\mu_V$) and
magnetic-quadrupole ($\mathcal{Q}_V$) moments~\cite{Hecht:1997uj}. In the present work, as we verified, since the quadrupole-moment contributions are practically negligible, we take into account only the monopole- and dipole-moment contributions for the $K^*$ exchange for brevity~\cite{Hawes:1998bz}, and the numerical inputs for the dipole moment are given as $\mu_{K^{*+}}=2.5$~\cite{Hecht:1997uj,Hawes:1998bz}.

Beside the EM-interaction  vertices defined in Eq.~(\ref{eq:LAGEM}), we are in a position to define the Lagrangians for the strong-interaction vertices which read:
%EQUATION>>>
\begin{eqnarray}
\label{eq:LAGST}
\mathcal{L}_{VNY}&=&-\bar{Y}\left[g_{VNY}
\rlap{/}{V}^*-\frac{g^T_{VNY}}{2(M_N+M_{Y})}
(\sigma\cdot G)\right]N+\mathrm{h.c.},
\cr
%-----------------------
\mathcal{L}_{PNY}&=&-ig_{PNY}
\bar{Y}\gamma_5PN+\mathrm{h.c.},
\cr
%-----------------------
\mathcal{L}_{SNY}&=&-g_{SNY}
\bar{Y}SN+\mathrm{h.c.},
\cr
%-----------------------
\mathcal{L}_{\gamma VNY}&=&-\frac{e_{V}g^T_{VNY}}
{2(M_N+M_{Y})}\bar{Y}(\sigma\cdot G)N+\mathrm{h.c.},
%-----------------------
\end{eqnarray}
%EQUAITON<<<
where $N$ and $Y$ indicate the ground-state nucleon and ground-state hyperon
fields. The relevant strong couplings can be estimated by the
potential models, such as the Nijmegen soft-core (NSC) potential model.
Following the NSC97a (NSC97f) model~\cite{Stoks:1999bz}, we have
$g_{KN\Lambda}=-13.42\,(-17.42)$, $g^V_{K^*N\Lambda}=-4.26\,(-6.11)$,
$g^T_{K^*N\Lambda}=-11.31\,(-14.85)$, and $g_{\kappa N\Lambda}=-8.31\,(-10.01)$. In the numerical calculations, we will consider these two parameter sets in order to verify the theoretical uncertainties. We define the magnetic-transition vertices for the hyperon resonance, $Y^*\equiv\Sigma^*(1385,3/2^+)$ contribution as follows:
%EQUATION>>>
\begin{eqnarray}
\label{eq:LAG} \mathcal{L}_{\gamma YY^*}&=&-\frac{ie \mu_{\gamma
YY^*}}{2M_{Y}} \bar{Y}^*_{\mu}\gamma_\nu \Gamma_5F^{\mu\nu} Y
-\frac{e \bar{\mu}_{\gamma YY^*}}{4M^2_{Y}}
\bar{Y}^{*}_{\mu}\Gamma_5F^{\mu\nu}(\partial_\nu
Y)+\mathrm{h.c.}\,, \cr
\mathcal{L}_{VNY^*}&=&-\frac{ig_{VNY^*}}{M_{V}}
\bar{Y}^*_\mu\gamma_\nu \Gamma_5G^{\mu\nu}N+\mathrm{h.c.}.
\end{eqnarray}
%EQUAITON<<<
As for $\Sigma^*$, we utilize the Rarita-Schwinger (RS) formalism for the spin-$3/2$~\cite{Rarita:1941mf,Read:1973ye}. In the above equations, the $4\times4$ matrix $\Gamma_5$ is chosen according to the parity of $Y^*$ as follows:
%EQUATION>>>
\begin{equation}
\label{eq:GAMMA}
\Gamma_5=\Big\{
\begin{array}{c}
\gamma_5\hspace{0.3cm}\mathrm{for}\hspace{0.3cm}\Pi(Y^*)=+1,\\
{\bf{1}}_{4\times4}\hspace{0.3cm}\mathrm{for}\hspace{0.3cm}\Pi(Y^*)=-1,
\end{array}
\end{equation}
%EQUAITON<<<
where $\Pi$ stands for the parity operator for $Y^*$. In the present work, we choose $\Pi=+1$, indicating $\Sigma^*$. For the input values of the relevant couplings in Eq.~(\ref{eq:LAG}), we use $\mu_{\gamma\Lambda\Sigma^{*0}}=3.78$, $\bar{\mu}_{\gamma\Lambda\Sigma^{*0}}=3.18$, and $g_{K^*N\Sigma^*}=-2.6$ by considering the experimental and theoretical studies as given in Ref.~\cite{Oh:2006hm} and references therein. All the relevant numerical inputs are summarized in Table~\ref{TABLE1}.
%TABLE>>>
\begin{table}[b]
\begin{tabular}{c|c|c|c||c|c|c|c||c|c||c}
\,\,\,\,$-g_{KN\Lambda}$\,\,\,\,
&\,\,\,\,$-g^V_{K^*N\Lambda}$\,\,\,\,
&\,\,\,\,$-g^T_{K^*N\Lambda}$\,\,\,\,
&\,\,\,\,$-g_{\kappa N\Lambda}$\,\,\,\,
&\,\,\,\,$g^\mathrm{charged}_{\gamma KK^*}$\,\,\,\,
&\,\,\,\,$g^\mathrm{neutral}_{\gamma KK^*}$\,\,\,\,
&\,\,\,\,$g^\mathrm{charged}_{\gamma \kappa K^*}$\,\,\,\,
&\,\,\,\,$g^\mathrm{neutral}_{\gamma \kappa K^*}$\,\,\,\,
&\,\,\,\,$\kappa_{\Lambda}$\,\,\,\,
&\,\,\,\,$\kappa_{\Lambda\Sigma^0}$\,\,\,\,
&\,\,\,\,$\mu_V$\,\,\,\,\\
\hline
$13.42 (17.42)$&$4.26 (6.11)$&$11.31(14.85)$&$8.31(10.01)$
&$-\frac{0.254}{\mathrm{GeV}}$
&$-\frac{0.388}{\mathrm{GeV}}$
&$\frac{0.120}{\mathrm{GeV}}$
&$\frac{0.240}{\mathrm{GeV}}$
&$-0.61$
&$1.61$
&$2.50$
\end{tabular}
\caption{Relevant couplings for the ground-state-particle contributions in Eq.~(\ref{eq:AMP}).}
\label{TABLE1} 
\end{table}
%TABLE>>>

Now, we are in a position to consider the nucleon-resonance contribution. In Ref.~\cite{Kim:2011rm}, several nucleon resonances were taken into account for the $K^*$ photoproduction, and it turns out that $D_{13}(2080)$ plays a most crucial role in reproducing observables in the vicinity of the threshold region. Hence, in the present work, similarly, we employ that nucleon resonance in order to see the resonance effects for the present reaction process. The relevant EM and strong interaction Lagrangians can be written as follows:
%EQUATION>>>
\begin{eqnarray}
\label{eq:LAGEMRE}
\mathcal{L}_{\gamma  ND_{13}}
&=& -\frac{ieh_{1D_{13}} }{2M_N} \bar N  \gamma_\nu F^{\mu\nu}
N^*_\mu -\frac{eh_{2D_{13}} }{(2M_N)^2} \partial_\nu \bar N 
F^{\mu\nu} N^*_\mu  + \mathrm{h.c.},              
\cr
\mathcal{L}_{K^* D_{13} \Lambda }
&=& -\frac{ig_{1D_{13}} }{2M_N} \bar{\Lambda} \gamma_\nu 
K^{*\mu\nu} N^*_\mu -\frac{g_{2D_{13}} }{(2M_N)^2} \partial_\nu \bar
\Lambda   K^{*\mu\nu} N^*_\mu                                          
+\frac{g_{3D_{13}} }{(2M_N)^2} \bar \Lambda 
\partial_\nu  K^{*\mu\nu} N^*_\mu + \mathrm{h.c.},     
\end{eqnarray}
%EQUATION>>>
where $N^*\equiv D_{13}(2080,3/2^-)$ stands for the field for the nucleon resonances with a certain spin and parity. The EM coupling constants are determined by using the experimental data for the  helicity amplitudes~\cite{Nakamura:2010zzi} and the quark model predictions of Ref.~\cite{Caps92}, which gives $h_{1D_{13}}=0.61$ and $h_{2D_{13}} = -0.62$. The strong coupling constants in Eq.~(\ref{eq:LAGEMRE}) can be determined from the theoretical estimations for the partial-wave decay
amplitudes~\cite{CR98b}, 
%EQUATION>>>
\begin{equation}
\label{eq:GGG}
\Gamma_{N^*\to K^*\Lambda}=\sum_\ell|G(\ell)|^2.
\end{equation}
%EQUAITON<<<
Here, $\Gamma_{N^*\to K^*\Lambda}$ is the partial-decay width of $N^* \to K^* \Lambda$.
The values for the partial-wave coupling strengths $G(\ell)$ can be found, e.g., in Ref.~\cite{CR98b}. Since the purpose of the present work is to investigate the role of nucleon resonances in the low-energy regions, it must be a good approximation to take into account only the lowest partial-wave ($s$-wave) contribution. Hence, we consider only the term proportional to  $g_{1D_{13}}$ in Eq.~(\ref{eq:LAGEMRE}), employing only $G(\ell=0)$.  Then by using Eq.~(\ref{eq:GGG}) and the theoretical value from Ref.~\cite{CR98b}, we obtain $|g_{1D_{13}}|=1.59$~\cite{Kim:2011rm}. The signs of these strong coupling constants are determined by fitting the experimental data. All the relevant parameters for the resonance contributions are listed in Table.~\ref{TABLE2}.
%TABLE>>>
\begin{table}[b]
\begin{tabular}{c|c|c||c|c|c|c|c}
\,\,\,\,$\mu_{\gamma\Lambda\Sigma^{*0}}$\,\,\,\,
&\,\,\,\,$\bar{\mu}_{\gamma\Lambda\Sigma^{*0}}$\,\,\,\,
&\,\,\,\,$g_{K^*N\Sigma^*}$\,\,\,\,
&\,\,\,\,$h_{1D_{13}}$\,\,\,\,
&\,\,\,\,$h_{2D_{13}}$\,\,\,\,
&\,\,\,\,$|g_{1D_{13}}|$\,\,\,\,
&\,\,\,\,$g_{2D_{13}}$\,\,\,\,
&\,\,\,\,$g_{3D_{13}}$\,\,\,\,\\
\hline
$3.78$&$3.18$&$-2.60$&$0.61$&$-0.62$
&$1.59$&$0.0$&$0.0$
\end{tabular}
\caption{Relevant couplings for the resonance-particle contributions in Eq.~(\ref{eq:AMP}).}
\label{TABLE2} 
\end{table}
%TABLE>>>

Taking into  account all the ingredients discussed so far, one can easily derive the the scattering amplitudes for the present production process:
%EQUATION>>>
\begin{eqnarray}
\label{eq:AMP}
i\mathcal{M}_s&=&
-i\bar{u}_2\left[g_{VNY}\rlap{/}{\epsilon}_2+\frac{g^T_{VNY}}{2(M_N+M_{Y})}
\left[\rlap{/}{k}_2,\rlap{/}{\epsilon}_2 \right]\right]
\left[\frac{\rlap{/}{k}_1F_s+(\rlap{/}{p}_1+M_N)F_{c}}{s-M^2_N} \right]
\left[e_N\rlap{/}{\epsilon}_1+\frac{e \kappa_N}{4M_N}
\left[\rlap{/}{\epsilon}_1,\rlap{/}{k}_1 \right]\right]u_1,
\cr
%-----------------
i\mathcal{M}_{u_Y}&=&
-i\bar{u}_2\left[e_{Y}\rlap{/}{\epsilon}_1+\frac{e \kappa_{Y}}
{4M_{Y}}\left[ \rlap{/}{\epsilon}_1,\rlap{/}{k}_1 \right]\right]
\left[\frac{(\rlap{/}{p}_2+M_{Y})F_{c}-\rlap{/}{k}_1F_u}{u-M^2_{Y}} \right]
\left[g_{VNY}\rlap{/}{\epsilon}_2+\frac{g^T_{VNY}}{2(M_N+M_{Y})}
\left[\rlap{/}{k}_2,\rlap{/}{\epsilon}_2 \right]\right]u_1,
\cr
%-----------------
i\mathcal{M}_{t_S}&=&-ig_{\gamma SV}g_{SNY}
\bar{u}_2\mathcal{D}_{t_S} u_1 \left[(k_1\cdot
k_2)(\epsilon_1\cdot\epsilon_2)
-(k_1\cdot\epsilon_2)(k_2\cdot\epsilon_1) \right]F_{t_S}, \cr
%-----------------
i\mathcal{M}_{t_P}&=&-g_{\gamma PV}g_{PNY}\bar{u}_2\gamma_5
\mathcal{D}_{t_P}u_1
\left(\epsilon_{\mu\nu\alpha\beta}k^\mu_1\epsilon^\nu
k^\alpha_2\eta^\beta\right) F_{t_P}, \cr
%-----------------
i\mathcal{M}_{t_V}&=&-ie_{V}(k_1\cdot\epsilon_2)\bar{u}_2
\mathcal{D}_{t_V}
\left[g_{VNY}\rlap{/}{\epsilon}_1+\frac{g^T_{VNY}}{2(M_N+M_{Y})}
\left[\rlap{\,/}{Q},\rlap{/}{\epsilon}_1 \right]
-\frac{g_{VNY}}{M^2_{V}}\rlap{\,/}{Q}(Q\cdot\epsilon_1)
\right]u_1F_c \cr &-&2ie_{V}(k_2\cdot\epsilon_1)\bar{u}_2
\mathcal{D}_{t_V}
\left[g_{VNY}\rlap{/}{\epsilon}_2+\frac{g^T_{VNY}}{2(M_N+M_{Y})}
\left[\rlap{\,/}{Q},\rlap{/}{\epsilon}_2 \right]
-\frac{g_{VNY}}{M^2_{V}}\rlap{\,/}{Q}(Q\cdot\epsilon_2)
\right]u_1F_c \cr &+&ie_{V}(\epsilon_1\cdot\epsilon_2)\bar{u}_2
\mathcal{D}_{t_V}
\left[g_{VNY}\rlap{/}{k}_2+\frac{g^T_{VNY}}{2(M_N+M_{Y})}
\left[\rlap{\,/}{Q},\rlap{/}{k}_2 \right]
-\frac{g_{VNY}}{M^2_{V}}\rlap{\,/}{Q}(Q\cdot k_2) \right]u_1F_c
\cr
%-----------------
&-&ie g_{VNY}
\mu_V\bar{u}_2\mathcal{D}_{t_V}\left\{\left[\rlap{/}{k}_1(\epsilon_1\cdot\epsilon_2)
-\rlap{/}{\epsilon}_1(k_1\cdot\epsilon_2)\right]
+\frac{\rlap{\,/}{Q}}{M^2_V}\left[(\epsilon_1\cdot Q)(k_1\cdot\epsilon_2)
-(k_1\cdot Q)(\epsilon_1\cdot\epsilon_2)\right\}
 \right]u_1F_{t_V},
%-----------------
\cr i\mathcal{M}_c&=&ie_{V} \bar{u}_2\left[\frac{g^T_{VNY}}
{2(M_N+M_{Y})} \left[\rlap{/}{\epsilon}_1,\rlap{/}{\epsilon}_2
\right] \right] u_1F_c,
%-----------------
\cr
i\mathcal{M}_{u_{Y^*}}&=& \frac{ie \mu_{\gamma
YY^*}g_{VNY^*}}{2M_VM_Y}
\bar{u}_2\Gamma_5\mathrm{M}_{4\times4}\Gamma_5u_1F_{u_{Y^*}} + \frac{ie
\bar{\mu}_{\gamma YY^*}g_{VNY^*}}{4M_VM^2_Y}
\bar{u}_2\Gamma_5\bar{\mathrm{M}}_{4\times4}\Gamma_5u_1F_{u_{Y^*}},
%-----------------
\cr
i\mathcal{M}_{s_{N^*}}  
&=&\frac{ig_{1D_{13}}}{2M_N}\epsilon^*_{2\nu} \bar{u}_2
\gamma_\sigma( k_2^\beta g^{\nu \sigma} - k_2^\sigma g^{\nu \beta})
\mathcal{D}_{N^*,\beta\alpha}                                                  
\left[ \frac{eh_{1D_{13}}}{2M_N} \gamma_\delta+
\frac{eh_{2D_{13}}}{4M_N^2} p_{1 \delta}\right]
(k_1^\alpha g^{\mu \delta} - k_1^\delta g^{\alpha\mu})\epsilon_{1\mu} u_1 F_{s_{N^*}}.                        
\end{eqnarray}
%EQUAITON<<<
Note that,  in deriving $\mathcal{M}_{t_V}$ and $\mathcal{M}_\mathrm{contact}$, we have used the minimal gauge substitution $\partial_\mu\to \partial_\mu+iV_\mu$ for the $K^*$
momentum derivative, and also assigned momentum $Q$ as $Q=k_2-k_1$. The relevant Lorentz structures $\mathrm{M}_{4\times4}$ and $\bar{\mathrm{M}}_{4\times4}$ for the
$Y^*$-pole contribution are given in Appendix. From the RS formalism~\cite{Rarita:1941mf}, the spin-$3/2$ fermion propagator reads:
%EQUATION>>>
\begin{equation}
\label{eq:DD}
\mathcal{D}^{\mu\nu}_{Y^*,N^*}(\ell)
=\frac{\rlap{/}{\ell}+M_{Y^*,N^*}}{\ell^2-M^2_{Y^*,N^*}}
\left[g^{\mu\nu}-\frac{1}{3}\gamma^\mu\gamma^\nu
-\frac{1}{3M_{Y^*,N^*}}(\gamma^\mu \ell^{\nu}-\gamma^\nu \ell^{\mu})
-\frac{2}{3M^2_{Y^*,N^*}}\ell^{\mu}\ell^{\nu}\right].
\end{equation}
%EQUAITON<<<
The mesonic propagators are simply defined as follows:
%EQUATION>>>
\begin{equation}
\label{eq:DDDD}
\mathcal{D}_{t_S}=\frac{1}{t-M^2_{t_S}},\,\,\,\,
\mathcal{D}_{t_P}=\frac{1}{t-M^2_{t_P}},\,\,\,\,
\mathcal{D}_{t_V}=\frac{1}{t-M^2_{t_V}}.
\end{equation}
%EQUAITON<<<
Here,  we used the notation $M_{t_{(S,P,V)}}$, indicating the mass of the (scalar, pseudoscalar, vector) particle, exchanged in the $t$ channel. Since the nucleon resonance has a finite full-decay width, we consider it by modifying the denominator in Eq.~(\ref{eq:DD}) as 
%EQUATION>>>
\begin{equation}
\label{eq:DENO}
\frac{1}{\ell^2-M^2_{D_{13}}}
\to\frac{1}{\ell^2-M^2_{D_{13}}+i\Gamma_{D_{13}}M_{D_{13}}},
\end{equation}
%EQUAITON<<<
where we have chosen $\Gamma_{D_{13}}\approx350$ MeV, which is almost an average value for those listed in Ref.~\cite{Nakamura:2010zzi}.

In order to consider the spatial distribution of the hadrons involved here, it is necessary to employ phenomenological form factors for the EM and strong vertices. In the present work, following the gauge-invariance form-factor prescription, suggested and applied in Refs.~\cite{Haberzettl:1998eq,Davidson:2001rk,Haberzettl:2006bn}, we make use of that defined as:
%EQUATION>>>
\begin{equation}
\label{eq:FF} F_{X}=\frac{\Lambda^4}{\Lambda^4+(X-M^2_{X})^2}\,,
\,\,\,\,F_c=1-(1-F_{s_N})(1-F_{t_V}),
\end{equation}
%EQUAITON<<<
where $X=s_{N,N^*},t_{S,P,V},u_{Y,Y^*}$, collectively. For instance, in the $t$ channel exchange, $X=t_{S,P,V}$, we have three different form factors, $F_{t_{S,P,V}}$, corresponding to each exchanged meson, ($\kappa,K,K^*$). Note that $F_X$ in Eq.~(\ref{eq:FF}) fulfills the
crossing symmetry. $F_c$ is a common form factor, which is taken into account for the gauge invariance of the scattering amplitude~\cite{Haberzettl:1998eq,Davidson:2001rk,Haberzettl:2006bn}. Moreover, this prescription satisfies the on-shell condition of form factors, i.e. $F(Q^2=0)=1$. The cutoff mass $\Lambda$ for the form factor will be determined to reproduce experimental data in the next Section.

Now, we want  to discuss on the $t$-channel Regge trajectories, which are essential to extend the low-energy Born  approximation to a certain high-energy region.  Moreover, as discussed in Refs.~\cite{Nam:2010au,Nam:2011np}, there can be a smooth interpolation between the conventional Born terms and Regge approaches even in the lower-energy region, although the Regge contribution becomes more significant in the higher-energy region. As in Refs.~\cite{Regge:1959mz,Vanderhaeghen:1997ts,Corthals:2005ce,Corthals:2006nz,Ozaki:2009wp},
the $t$-channel Regge propagator can be written generically as follows:
%EQUAITON<<<
\begin{equation}
\label{eq:REGGE} \mathcal{D}^{\mathrm{Regge}}_T=
\frac{\pi\alpha'_T}{\sin(\pi\alpha_T)\Gamma(1+\alpha_T-n_T)}
\left(\frac{s}{s_0} \right)^{\alpha_T-n_T},
\end{equation}
%EQUAITON<<<
where $T=(t_S,t_P,t_V)$ and $(n_{t_S},n_{t_P},n_{t_V})=(0,0,1)$. $\alpha_T$ denotes the Regge trajectory for the strange meson, $(\kappa,K,K^*)$ as a function of $t$ with the slope of the trajectories, $\alpha'_T$. As for each meson, the
trajectory reads:
%EQUATION>>>
\begin{equation}
\label{eq:ALPHA}
\alpha_{t_S}(t)=\alpha'_{t_S}(t-M^2_{t_S}),\,\,\,\,
\alpha_{t_P}(t)=\alpha'_{t_P}(t-M^2_{t_P}),\,\,\,\,
\alpha_{t_V}(t)=\alpha'_{t_V}(t-M^2_{t_V})+1,
\end{equation}
%EQUAITON<<<
where the slope parameters are chosen to  be $(\alpha'_{t_S},\alpha'_{t_P},\alpha'_{t_V})=(0.7,0.7,0.85)/\mathrm{GeV}^2$. These slope parameters are taken from Refs.~\cite{Vanderhaeghen:1997ts,Corthals:2005ce,Corthals:2006nz,Ozaki:2009wp}, and we will keep these values unchanged throughout this work. We set $s_{0}=1.0$ GeV for all the mesons conventionally~\cite{Vanderhaeghen:1997ts,Corthals:2005ce,Corthals:2006nz}. Here is a caveat: In deriving Eq.~(\ref{eq:REGGE}), all the even and odd spin trajectories are assumed to be degenerate, although in reality these trajectories are not degenerated~\cite{Vanderhaeghen:1997ts,Corthals:2005ce,Corthals:2006nz}. Moreover, for convenience, we have set the phase factor for the propagators to be positive unity as done in Ref.~\cite{Ozaki:2009wp}. Hereafter, we will use a notation $i\mathcal{M}^{\mathrm{Regge}}$ for the amplitude thus constructed with the Regge propagators in Eq.~(\ref{eq:REGGE}) as
%EQUATION>>>
\begin{equation}
\label{eq:DDDDD} \mathcal{D}_{(t_S,t_P,t_V)}
\to[t-M_{(t_S,t_P,t_V)}]\mathcal{D}^{\mathrm{Regge}}_{(t_S,t_P,t_V)}.
\end{equation}
%EQUAITON<<<
Note  that this simple replacement causes a problem to the gauge invariance of the scattering amplitude. Hence, we follow the remedy as suggested in Ref.~\cite{Ozaki:2009wp}, in which one multiplies $(t-M^2_{t_V})\mathcal{D}^\mathrm{Regge}_{t_V}$ to the nonzero terms in
$ik_1\cdot\mathcal{M}_\mathrm{total}=ik_\gamma\cdot\mathcal{M}_\mathrm{total}$, by setting all the form factors unity.

As mentioned briefly above, we consider the interpolation between the Feynman and Regge propagators. It is worth mentioning that this Feynman-Regge interpolation is highly phenomenological without concrete theoretical bases. However, in a practical point of view, this prescription nevertheless reproduces data qualitatively well as demonstrated in Refs.~\cite{Nam:2010au,Nam:2011np}. Hence, we want to employ the prescription  in a simplified form in the present work, and to see its reliability. Analytically, this interpolation can be achieved by the following replacements:
%EQUATION>>>
\begin{eqnarray}
\label{eq:PARA}
F_{(t_S,t_P,t_V)}&\to&\mathcal{R}_{(S,P,V)}[t-M^2_{(t_S,t_P,t_V)}]
\mathcal{D}^{\mathrm{Regge}}_{(t_S,t_P,t_V)}
+(1-\mathcal{R}_{(S,P,V)})F_{(t_S,t_P,t_V)}\equiv
\bar{F}_{(t_S,t_P,t_V)}, \cr
F_c&\to&\mathcal{R}_c(t-M^2_{t_V})\mathcal{D}^{\mathrm{Regge}}_V
+(1-\mathcal{R}_c)F_c\equiv \bar{F}_{c},
\end{eqnarray}
%EQUAITON<<<
where $\bar{F}_{(t_S,t_P,t_V)}$ and $\bar{F}_{c}$ denote the form factors to interpolate the propagators between the Feynman and Regge regimes. As easily understood in Eq.~(\ref{eq:PARA}), one can mix the pure Feynman and Regge propagators by adjusting the
values of the mixing parameter, $\mathcal{R}_{S,P,V,c}$, from zero to unity. Setting $\mathcal{R}_{S,P,V,c}=0$ and $1$, we recover the Feynman and Regge ones, respectively. The values for those $\mathcal{R}_{S,P,V,c}$ will be determined to reproduce the experimental data in the next Section. It is worth noting that, in principle, the mixing parameter can be a function of $s$ as well as $t$ as discussed in Refs.~\cite{Nam:2010au,Nam:2011np}. However, in this work, we treat the mixing parameters to vary from zero to unity for convenience. Even in this simplified version, one can figure out what portion of the Regge contributions are in each energy region. Finally, the total scattering amplitude with the interpolation prescription on top of the gauge invariance can be written as follows:
%EQUATION>>>
\begin{eqnarray}
\label{eq:MMM}
\mathcal{M}_\mathrm{total}=
(\mathcal{M}^E_s+\mathcal{M}^E_{t_V}+\mathcal{M}_c)\bar{F}_c
+(\mathcal{M}^M_{u_\Lambda}+\mathcal{M}_{u_\Sigma}+\mathcal{M}_{u_{\Sigma^*}})F_u
+\mathcal{M}_{s_{N^*}}F_{s}+\mathcal{M}_{t_S}\bar{F}_{t_S}
+\mathcal{M}_{t_P}\bar{F}_{t_P}+\mathcal{M}^M_{t_V}\bar{F}_{t_V},
\end{eqnarray}
%EQUAITON<<<
where  the superscripts $E$ and $M$ indicate the amplitudes with electric and magnetic interaction vertices, respectively.  We note that the scheme for the form-factors as shown in Eq.~(\ref{eq:MMM}) is equivalent in principle to that in Ref.~\cite{Oh:2006hm}. Note that, in contrast, we employ the same type of the form factor defined in Eq.~(\ref{eq:FF}) for all the channels with a single cutoff mass $\Lambda$, whereas the dipole-type form factors were used for the $t$-channel contributions with different cutoff masses in Ref.~\cite{Oh:2006hm}.

%--------------------------------------------------
\section{Numerical results and Discussion}
\begin{figure}[t]
\begin{tabular}{cc}
\includegraphics[width=8.5cm]{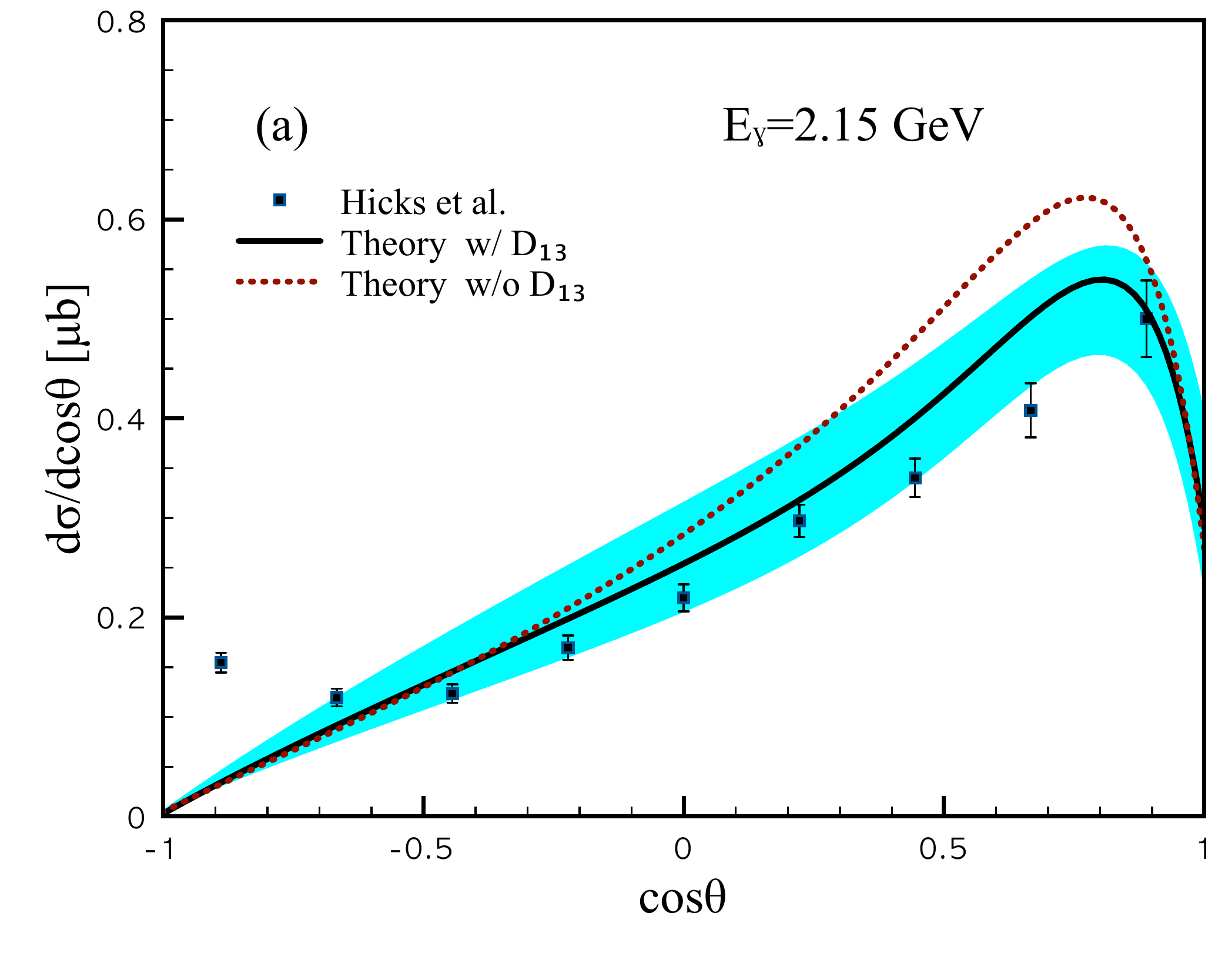}
\includegraphics[width=8.5cm]{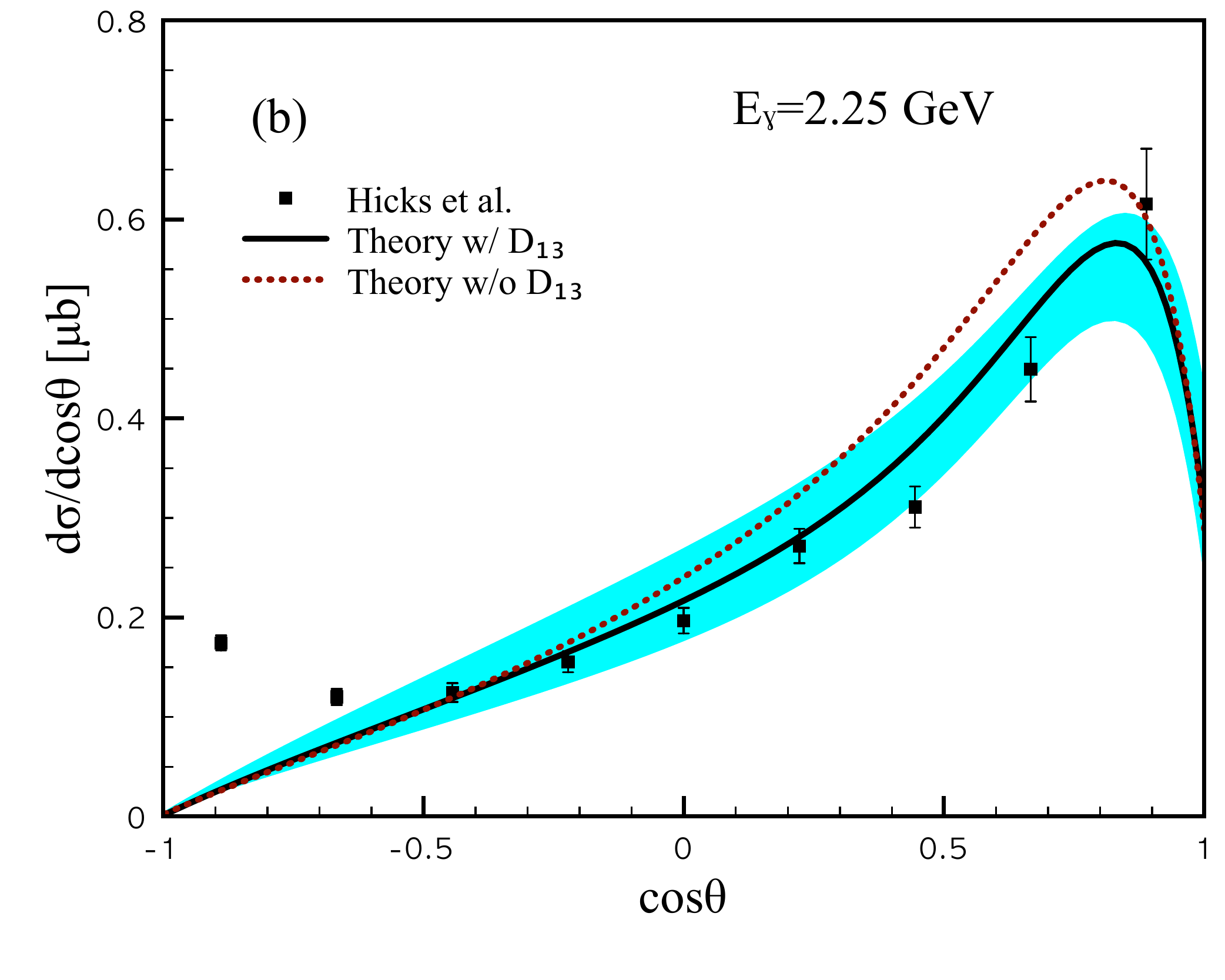}
\end{tabular}
\begin{tabular}{cc}
\includegraphics[width=8.5cm]{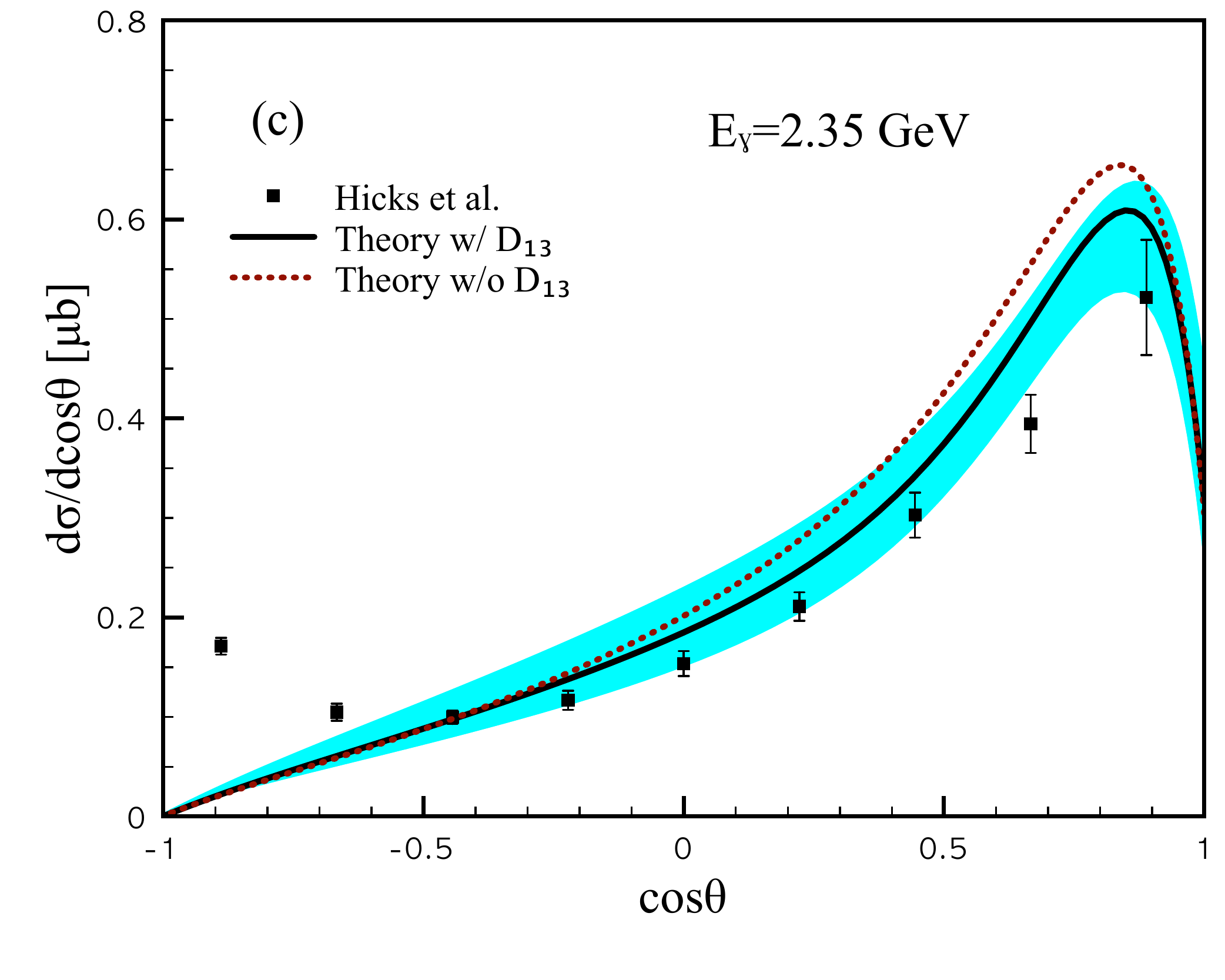}
\includegraphics[width=8.5cm]{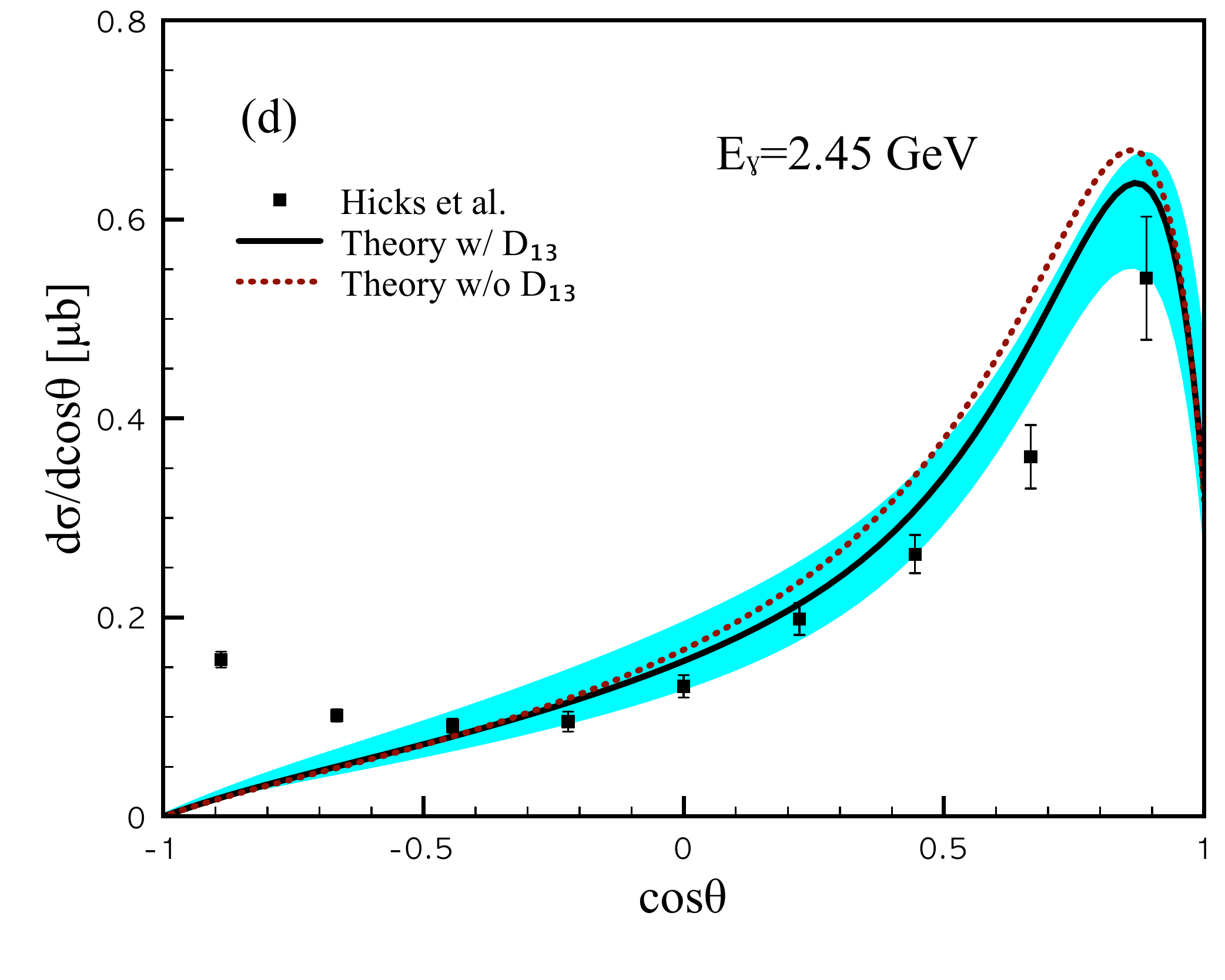}
\end{tabular}
\begin{tabular}{cc}
\includegraphics[width=8.5cm]{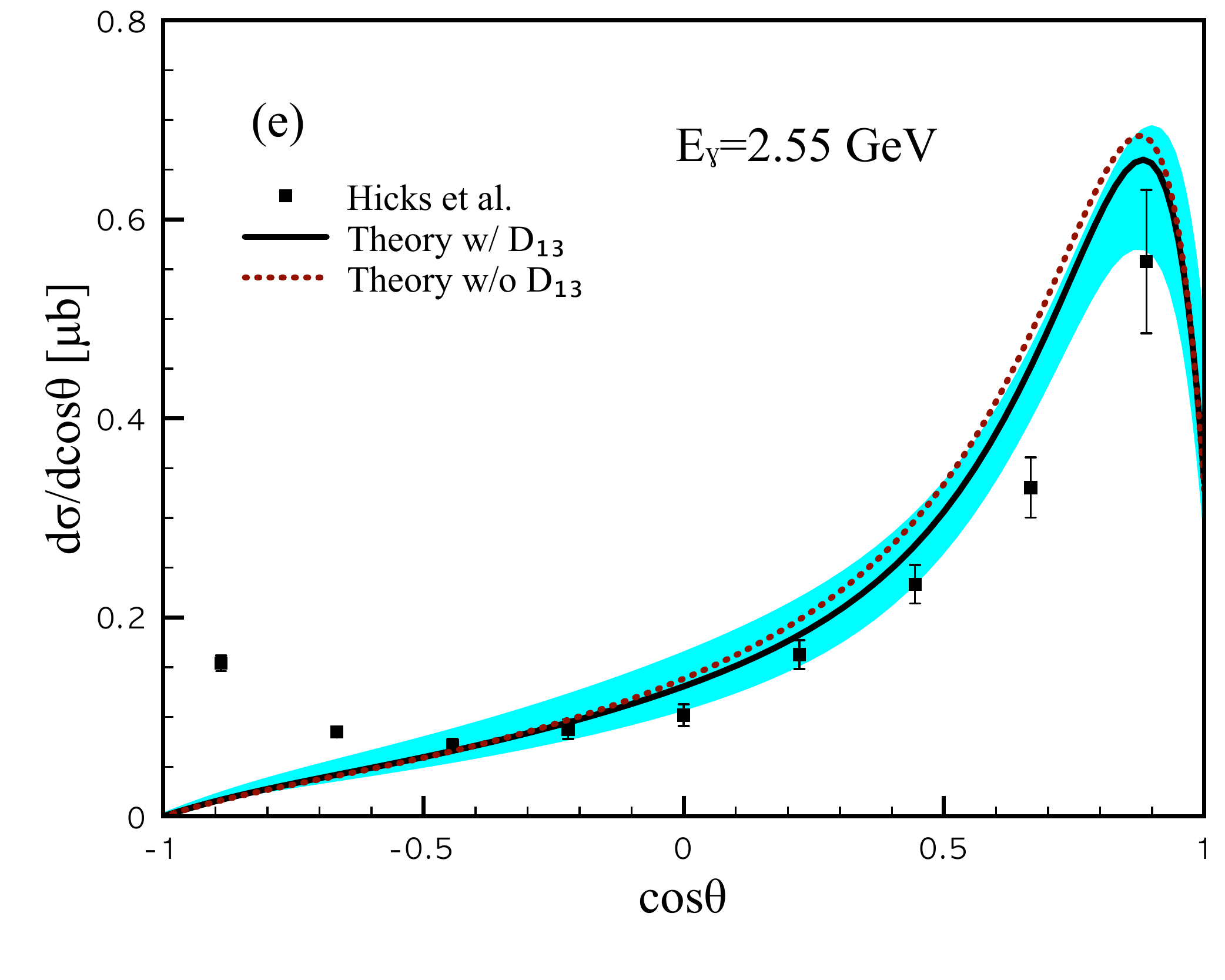}
\includegraphics[width=8.5cm]{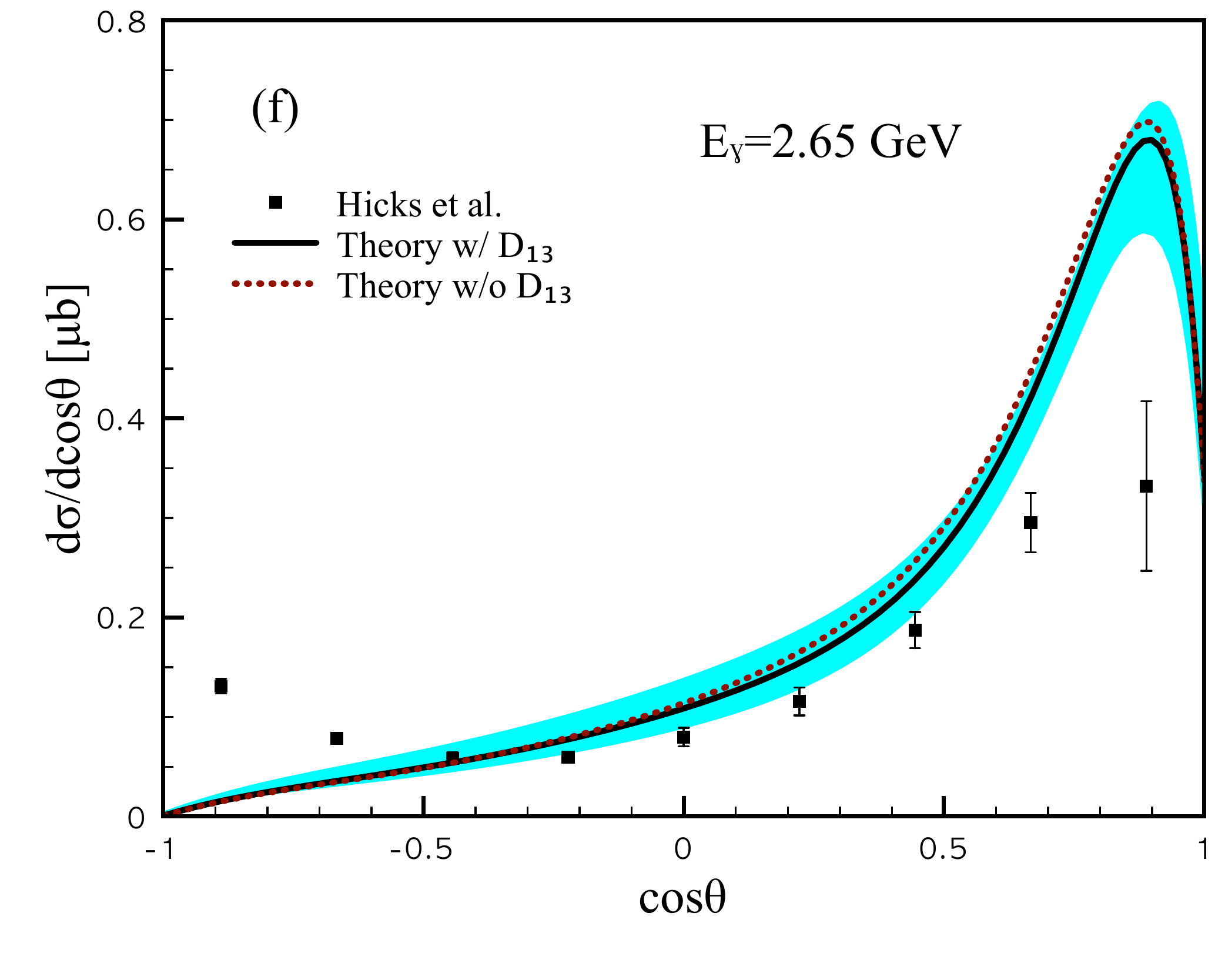}
\end{tabular}
\caption{(Color online) Differential cross sections as functions
of $\cos\theta$ for $E_\gamma=(2.15\sim2.65)$ GeV shown in the
panels (a $\sim$ f). Combined theoretical uncertainties are given by the shaded areas for $\Lambda=(500\sim550)$ MeV and $\mathcal{R}_{S,P,V}=(0.45\sim0.6)$, using NSC97a and NSC97f. The solid and dot lines indicate the numerical results for the averaged parameters in Eq.~(\ref{eq:AVE}) with and without $D_{13}$. The experimental data are taken from the CLAS experiment by Hicks {\it et al.}~\cite{Hicks:2010pg}.} 
\label{FIG1}
\end{figure}

%--------------------------------------------------
In this Section, we present the numerical results and related discussions for the $K^*\Lambda$ photoproduction off the proton target. First, we present the differential cross section  in the
panels (a $\sim$ f) of Figure.~\ref{FIG1} as a function of $\cos\theta$, in which $\theta$ stands for an angle between the incident photon and outgoing kaon in the center-of-mass (cm) frame, for the different photon energies, $E_\gamma=(2.16\sim2.65)$ GeV. The preliminary experimental data are taken from Ref.~\cite{Hicks:2010pg}. As we mentioned in the previous Section, our numerical results are computed using the interpolation, given by Eq.~(\ref{eq:MMM}), and all the model parameters, such as the cutoff mass $\Lambda$ in Eq.~(\ref{eq:FF}) and mixing parameter $\mathcal{R}$ in Eq.~(\ref{eq:PARA}), are determined so as to reproduce the experimental data. In order to test the theoretical uncertainties arise from those various parameters, we vary them as follows:
%EQUATION>>>
\begin{equation}
\label{eq:PARAVARY}
\Lambda=(500\sim550)\,\mathrm{MeV},\,\,\,\,\mathcal{R}_{S,P,V}=(0.45\sim0.6),
\end{equation}
%EQUAITON<<<
for two different sets for the strong couplings from NSC97a and NSC97f as in Table.~\ref{TABLE1}. For convenience, we fix $\mathcal{R}_c=0.0$ as a trial. For instance, the value $\mathcal{R}_{S,P,V}=0.5$ indicates physically that the production strength is resulted evenly by the Regge and Feynman contributions in the energy region $E_\gamma=(2.15\sim2.65)$ GeV, whereas all the electric contributions are those from the pure Born terms as understood by $\mathcal{R}_c=0$ in Eqs.~(\ref{eq:PARA}) and (\ref{eq:MMM}). The combined theoretical uncertainties from $\Lambda$, $\mathcal{R}_{S,P,V}$, and the coupling sets are presented by the shaded bands in Figure.~\ref{FIG1}. The solid and dot lines indicate the numerical results with and without $D_{13}$, respectively, using the averaged parameters, which are
%EQUATION>>>
\begin{equation}
\label{eq:AVE}
\Lambda=525\,\mathrm{MeV},\,\,
\mathcal{R}_{S,P,V}=0.53,\,\,
g_{KN\Lambda}=-15.42,\,\,
g^V_{K^*N\Lambda}=-5.19,\,\,
g^T_{K^*N\Lambda}=-13.08,\,\,
g_{\kappa N\Lambda}=-9.16.
\end{equation}
%EQUAITON<<<

Since the data are reproduced quantitatively well for all the photon energies as shown in the figure, we may consider that the interpolation prescription is very useful to describe the production mechanism for the wide energy range, simultaneously. There appears, however, a sizable underestimate in the backward scattering region, since we do not have enough $u$-channel contributions from the hyperon poles, except for $\Lambda$, $\Sigma$, and $\Sigma^*$. Similar observation was already reported in Ref.~\cite{Kim:2011rm}. Although the inclusion of more hyperon in the $u$-channel may lead to better reproduction of the data, we did not do that, since it also brings more theoretical uncertainties, considering the lack of experimental and theoretical information for them. As for $E_\gamma=2.65$ GeV, we observe a strong
overshoot in the forward-scattering region in comparison to the experimental data. We verified that this originated from the strong interference between the $K$ and $\kappa$ contributions at this energy region. To tame such an overshoot, one may need destructive interferences with higher-spin meson exchanges, such as the tensor mesons, for instance. Another possible remedy for this overshoot is to make the parameter $\mathcal{R}$ a function of energy as in Ref.~\cite{Nam:2010au,Nam:2011np}. It turns out that the resonance contributions are not so significant for the differential cross sections as understood by seeing the curves with (solid) and without (dot) $D_{13}$. As the energy increases, the resonance contribution gets diminished as expected, i.e. the resonance is effective near the threshold with a finite decay width.

\begin{figure}[t]
\begin{tabular}{cc}
\includegraphics[width=8.5cm]{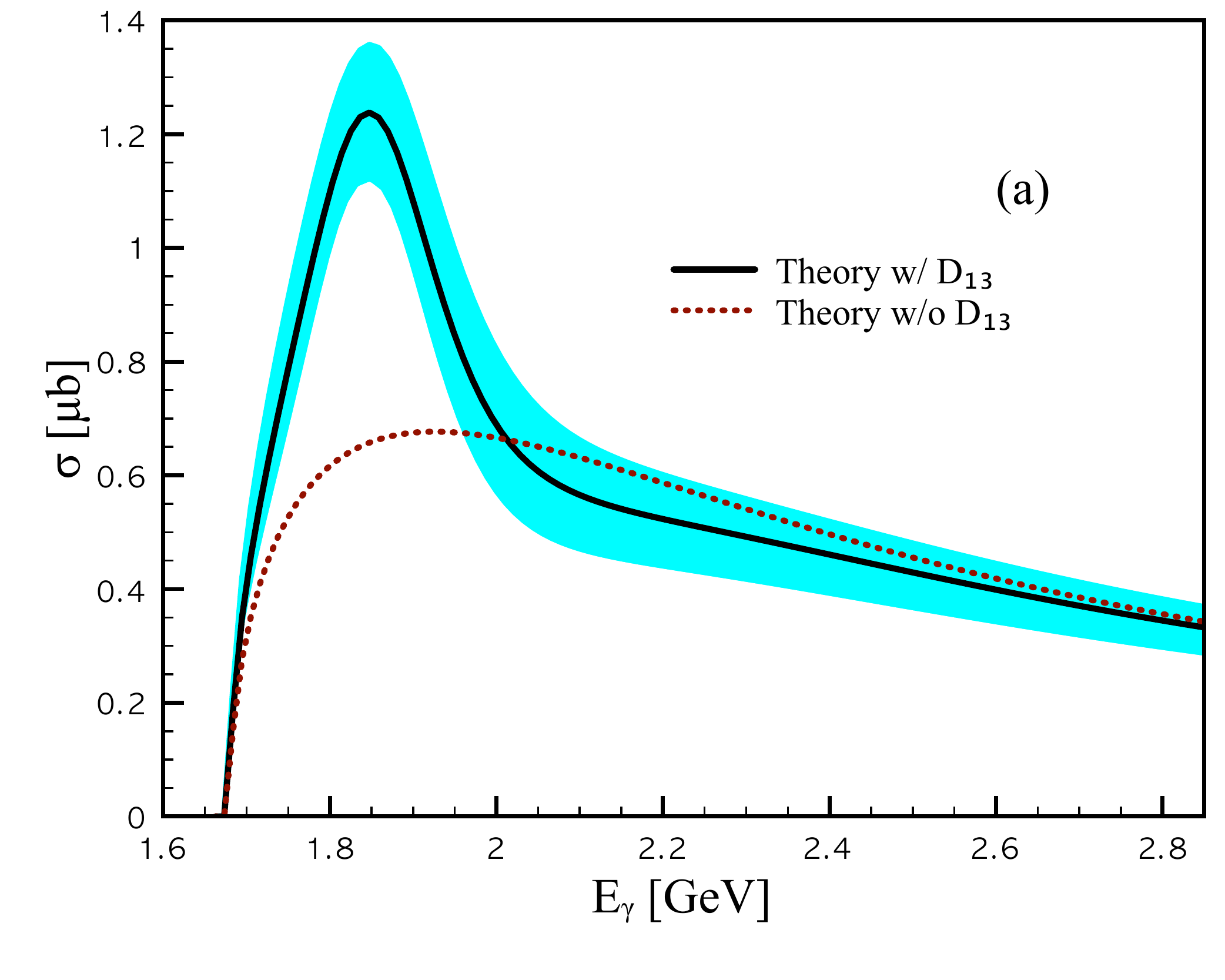}
\includegraphics[width=8.5cm]{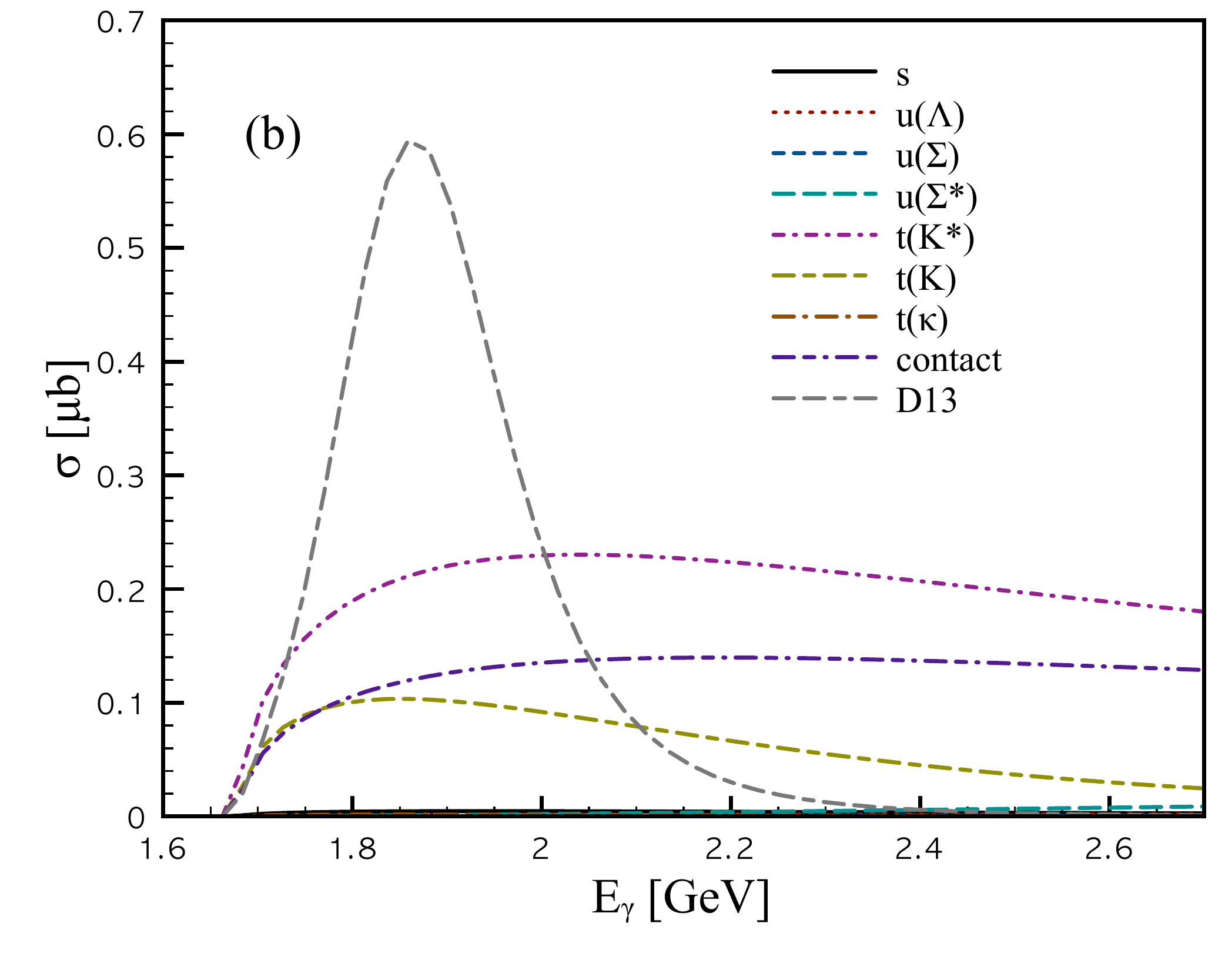}
\end{tabular}
\caption{(Color online) Total cross section for the
proton target in the panel (a) in the same manner with Figure~\ref{FIG1} using the averaged parameters in Eq.~(\ref{eq:AVE}). We also show each contribution separately in the panel (b).}
\label{FIG2}
\vspace{1cm}
\begin{tabular}{cc}
\includegraphics[width=8.5cm]{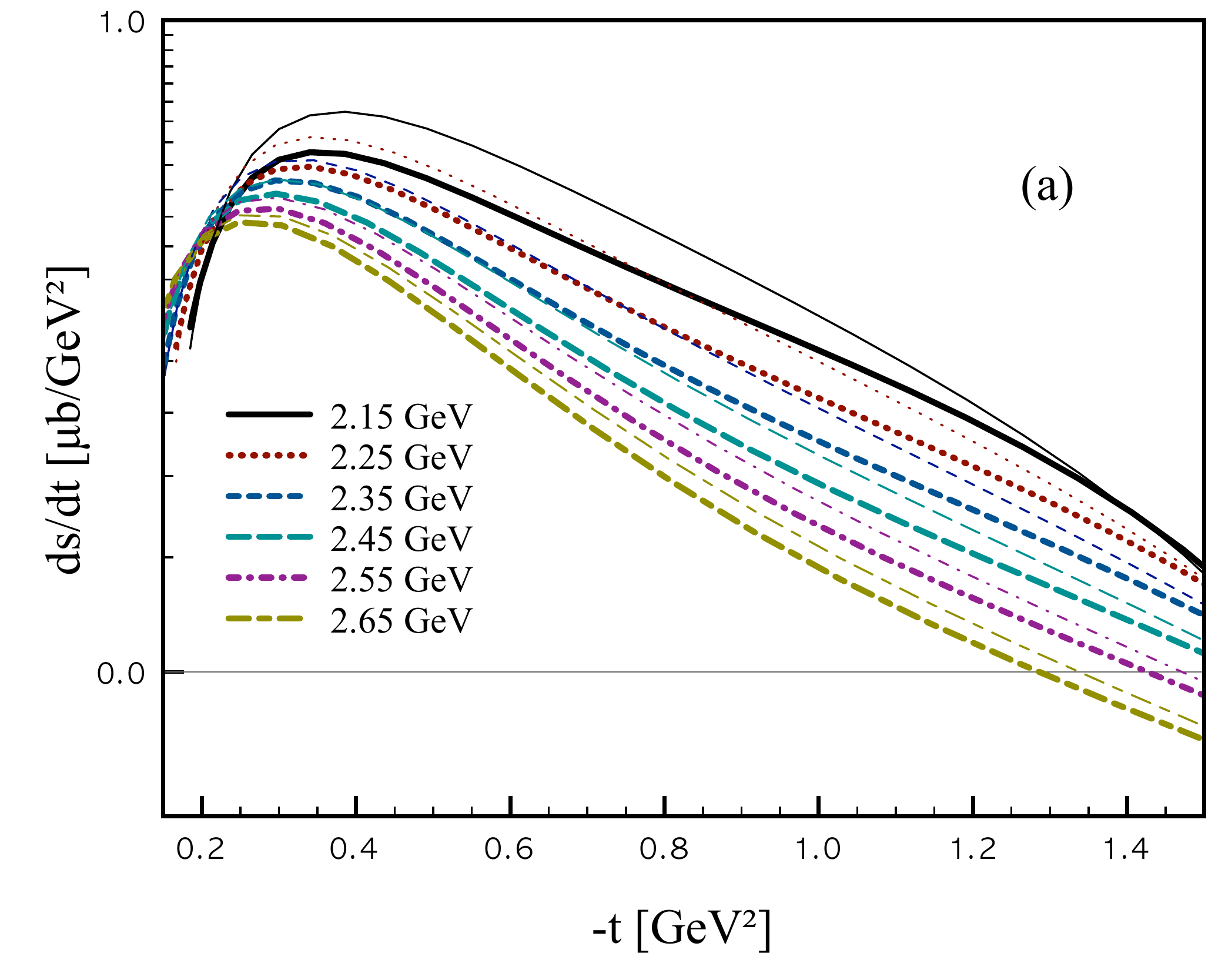}
\includegraphics[width=8.5cm]{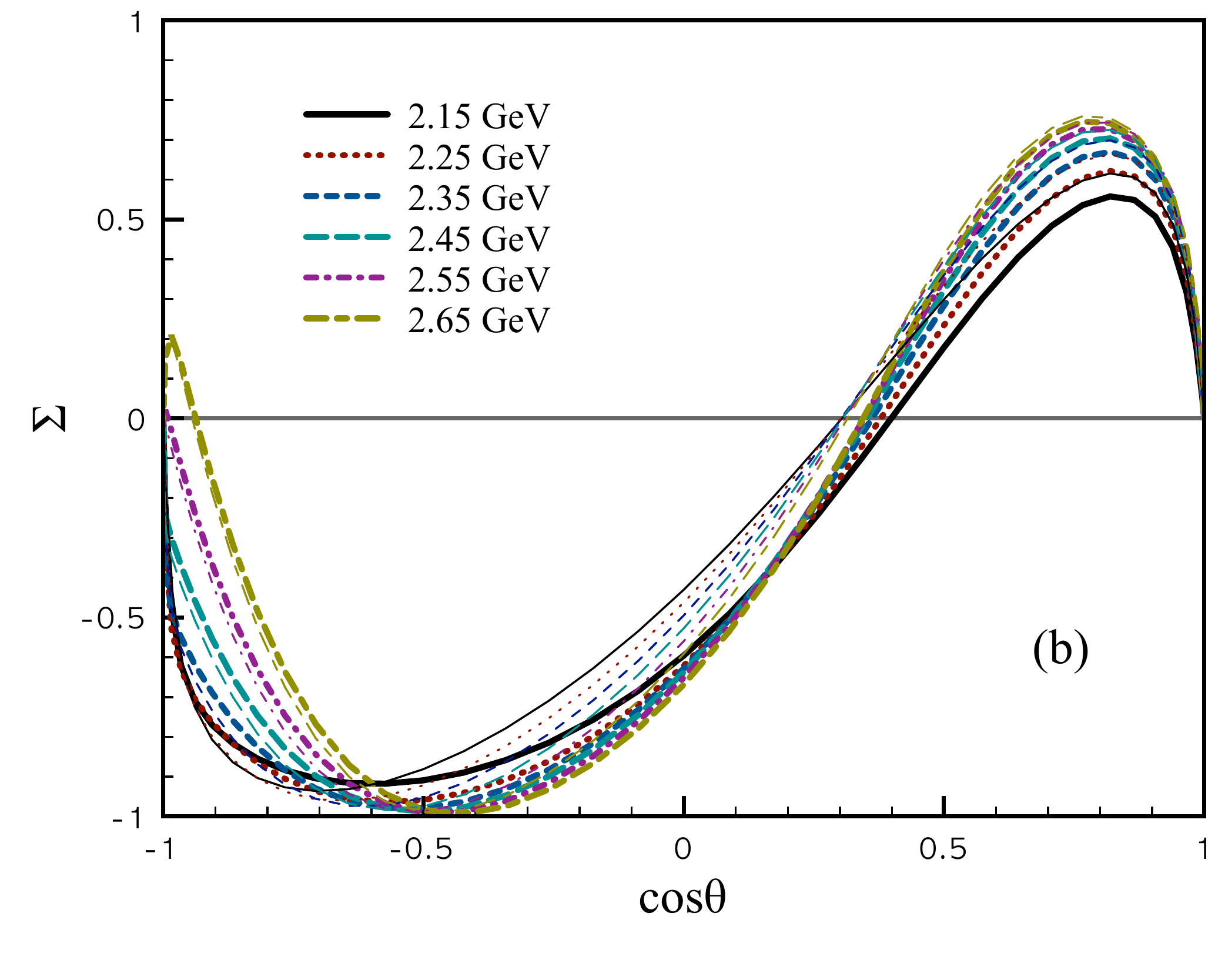}
\end{tabular}
\caption{(Color online) $t$-channel momentum-transfer dependence of $d\sigma/dt$ as a function of $-t$ for the proton target for various photon energies in the panel (a). Photon-beam asymmetry $\Sigma$ defined in Eq.~(\ref{eq:BA}) as a function of $\cos\theta$ in the panel (b). The thick and thins lines indicate the results with and without $D_{13}$, respectively. All the numerical results are computed using the averaged parameters in Eq.~(\ref{eq:AVE}).} 
\label{FIG3}
\end{figure}

In the same manner of presentation with Figure~\ref{FIG1}, we show the numerical results for the total cross section within the present framework in Figure~\ref{FIG2}. In the panel (a) of Figure~\ref{FIG2}, we draw the curves as functions of $E_\gamma$. If we compare the curves with (solid) and without (dot) $D_{13}$, there appears considerable enhancement near the threshold region due to the resonance. In the panel (b) of Figure~\ref{FIG2}, we draw each contribution separately to the present reaction process. Note that only $K$- and $K^*$-exchanges, $D_{13}$ and contact-term contributions are clearly visible, whereas other contributions are almost negligible in comparison to them. Therefore, we can conclude that the $(K,K^*)$-exchange and contact-term contributions play most dominant roles to produce the curves for the total cross section. Again, we clearly see the resonance contribution, peaking at $E_\gamma=(1.8\sim1.9)$ GeV. 

Fianally, we present the numerical results for the momentum-transfer dependence of the $d\sigma/dt$ as a function of $-t$ (a) and the photon-beam asymmetry $\Sigma$ defined in Eq.~(\ref{eq:BA}) as a function of $\cos\theta$ (b) in Figure~\ref{FIG3} for $E_\gamma=(2.15\sim2.65)$ GeV :
%EQUATION>>>
\begin{equation}
\label{eq:BA}
\Sigma=\frac{d\sigma_\parallel-d\sigma_\perp}{d\sigma_\parallel+d\sigma_\perp},
\end{equation}
%EQUAITON<<<
where $d\sigma_\parallel$ and $d\sigma_\perp$ represent  the differential cross sections with the polarized photon beam being parallel and perpendicular to the reaction plane, respectively. Note that all the curves were computed using the parameters given in Eq.~(\ref{eq:AVE}). The thick and thin lines represent the numerical results with and without $D_{13}$. We observe that the resonance contribution from $D_{13}$ is not so significant for those physical observables, as shown in Figure~\ref{FIG3}. The numerical results for the momentum-transfer dependence of the differential cross section and the beam asymmetry will be useful guides for ongoing and future experiments which can test the reliability of the present theoretical framework, in particular, for the prescription for the interpolation between the Feynman-Regge
propagators.

%--------------------------------------------------
\section{Summary and conclusion}
%--------------------------------------------------
In this letter,  we have studied the $K^*\Lambda$ photoproduction off the proton target, $\gamma p\to K^{*+}\Lambda$. For this purpose, we employed the effective Lagrangian approach at tree level. To consider the Regge trajectories in the strange-meson exchanges in the $t$ channel, we took into account the interpolation between the Feynman-Regge regimes
based on the phenomenological prescription. We included baryon-resonance contributions from $D_{13}(2080,3/2^-)$ as well as $\Sigma^*(1385,3/2^+)$ in the $s$ and $u$ channels, respectively. The gauge-invariant form-factors scheme was used consistently with all the channels, which satisfies the crossing symmetry as well as the on-shell condition, even in the presence of the Feynman-Regge interpolation prescription. The estimations on the theoretical uncertainties generated from various model parameters were also performed. Below, we list the important observations of the present theoretical study:
%ITEMIZE>>>
\begin{itemize}
%-------------------
\item The recent CLAS preliminary data for the angular dependence~\cite{Hicks:2010pg} are reproduced qualitatively well with the interpolation prescription for the relatively wide photon-energy regions, i.e. $E_\gamma=(2.15\sim2.65)$ GeV. From the numerical results, we find that the Regge contribution turns out to be effective even in the lower-energy region, whereas all the contributions from the electric components of the production amplitude are those from the pure Born terms. It is also found that the contribution from $D_{13}$ is not significant for the angular dependences and decreases as $E_\gamma$ increases. 
%-------------------
\item There appear sizable underestimates in the angular dependences at the backward
angles, which may be due to the insufficient effects from the $u$-channel hyperon resonances. We
also observe the overshoots in the forward-scattering region at high energy. In order to improve the model prediction at the forward-scattering region, one may need to consider the high-spin meson exchange for the destructive interferences with those considered here in the $t$ channel or more realistic mixing parameters $\mathcal{R}$.
%-------------------
\item Numerical results for the total cross section turn out to be effected much by the inclusion of the nucleon resonance $D_{13}$ in the vicinity of the threshold. From the numerical results, we also verified  that the $(K,K^*)$-exchange and contact-term contributions dominate the present reaction process in addition to the nucleon resonance near the treshold, while the $\kappa$ exchange plays a minor role. The momentum-transfer dependence and the photon-beam asymmetry are also estimated theoretically for the future experimental data analyses.
\end{itemize}
%ITEMIZE>>>

It is worth mentioning that, although the interpolation of the Feynman and Regge propagators plays a significant role to reproduce the data qualitatively well, we still have several uncertainties and unknown factors in this highly phenomenological approach. We also note the possibility that the prescription for the interpolation can be replaced by other production mechanisms such as various nucleon and hyperon resonances, and tested within the present framework. We also emphasize that the experimental data used in this work are still preliminary, so that our theoretical conclusions based on them can be modified by comparison with published data in the near future. More sophisticated works, including sufficient nucleon and hyperon resonances, realistic mixing parameter $\mathcal{R}$ as a function of energy, and heavier strange-meson contributions, are under progress and will appear elsewhere.
%--------------------------------------------------
\section*{Acknowledgment}
%--------------------------------------------------
The authors are grateful  to K.~Hicks for CLAS collaboration at Jefferson laboratory who kindly provided us the experimental data, and also appreciate his detailed and constructive comments on the present work. They also thank A.~Hosaka, Y.~Oh, and H.~-Ch.~Kim for fruitful discussions. The authors sincerely acknowledge that the technical supports for the present numerical calculations were given by S.~H.~Kim. This work was supported by the grant NRF-2010-0013279 from National Research Foundation (NRF) of Korea.

%--------------------------------------------------
\section*{Appendix}
%--------------------------------------------------
The relevant Lorentz structures  $\mathrm{M}_{4\times4}$ and
$\bar{\mathrm{M}}_{4\times4}$ in Eq.~(\ref{eq:AMP}) are defined as
follows:
%EQUATION>>>
\begin{eqnarray}
\label{eq:MAT} \mathrm{M}_{4\times4}
&\equiv&(k_{1,\mu}\rlap{/}{\epsilon}_1-\rlap{/}{k}_1\epsilon_{1,\mu})
\mathcal{D}^{\mu\nu}_{Y^*}(k_{2,\nu}\rlap{/}{\epsilon}_2-\rlap{/}{k}_2\epsilon_{2,\nu}),
\cr &=&-\frac{1}{3M^2_{Y^*}}\frac{\rlap{/}{q}_u-M_{Y^*}}
{u-M^2_{Y^*}}\left[
\rlap{/}{\epsilon}_1\rlap{/}{\epsilon}_2\mathcal{F}(k_1,k_2)
-\rlap{/}{\epsilon}_1\rlap{/}{k}_2\mathcal{F}(k_1,\epsilon_2)
+\rlap{/}{k}_1\rlap{/}{k}_2\mathcal{F}(\epsilon_1,\epsilon_2)
-\rlap{/}{k}_1\rlap{/}{\epsilon}_2\mathcal{F}(\epsilon_1,k_2)
 \right],
\cr \bar{\mathrm{M}}_{4\times4} &\equiv&[(\epsilon_1\cdot
p_2)k_{1,\mu}-(k_1\cdot p_2)\epsilon_{1,\mu}]
\mathcal{D}^{\mu\nu}_{Y^*}(k_{2,\nu}\rlap{/}{\epsilon}_2-\rlap{/}{k}_2\epsilon_{2,\nu}),
\cr &=&-\frac{1}{3M^2_{Y^*}}\frac{\rlap{/}{q}_u-M_{Y^*}}
{u-M^2_{Y^*}}\left[
\rlap{/}{\epsilon}_1\rlap{/}{\epsilon}_2\mathcal{F}(k_1,k_2)
-\rlap{/}{\epsilon}_1\rlap{/}{k}_2\mathcal{F}(k_1,\epsilon_2)
+\rlap{/}{k}_1\rlap{/}{k}_2\mathcal{F}(\epsilon_1,\epsilon_2)
-\rlap{/}{k}_1\rlap{/}{\epsilon}_2\mathcal{F}(\epsilon_1,k_2)
 \right],
\end{eqnarray}
%EQUAITON<<<
where $q_u\equiv p_2-k_1$ and the $4\times4$ matrix function $\mathcal{F}$ is defined as
%EQUATION>>>
\begin{equation}
\label{eq:} \mathcal{F}(k_a,k_b)= [\rlap{/}{k}_a
\rlap{/}{k}_b-3(k_a\cdot k_b)]M^2_{Y^*} +2(k_a\cdot q_u)(k_b\cdot
q_u).
\end{equation}
%EQUAITON<<<
%--------------------------------------------------

%--------------------------------------------------
%\end{document}

\newpage
%FIGURE>>>
%FIGURE<<<

%FIGURE>>>
%FIGURE<<<
%--------------------------------------------------
\end{document}